\documentclass[fp,twocolumn]{jpsj3}
\usepackage{graphicx}
\usepackage{bm}
\usepackage{amsmath, amssymb}

\usepackage{color}

\title{Spontaneously symmetry breaking states
in the attractive SU$(N)$ Hubbard model}

\author{Akihisa Koga and Hiromasa Yanatori}
\inst{Department of Physics, Tokyo Institute of Technology, 
Tokyo 152-8551, Japan}

\date{\today}%

\abst{
We investigate spontaneously symmetry breaking states in
the attractive SU($N$) Hubbard model at half filling.
Combining dynamical mean-field theory with the continuous-time 
quantum Monte Carlo method,
we obtain the finite temperature phase diagrams for the superfluid state.
When $N>2$, the second-order phase transition occurs 
in the weak coupling region, while
the first-order phase transition with the hysteresis appears 
in the strong coupling region.
We also discuss the stability of the density wave state and
clarify the component dependence of the maximum critical temperature.
}

\begin{document}

\maketitle

\section{Introduction}
Low temperature properties in correlated fermion systems 
have attracted much attention
since the discovery of the high-$T_c$ cuprates~\cite{Bednorz}.
In the strongly correlated electron systems, 
the spin degrees of freedom play a crucial role in 
stabilizing interesting phenomena at low temperatures such as
the magnetism, superconductivity, and Mott phase transitions.
The orbital degrees of freedom enrich the system, where 
the superconductivity and colossal magnetoregistance 
have been observed in the ruthenate~\cite{Maeno} and magnanites.~\cite{Tokura}.
Recently, an optical lattice system, where the periodic potential 
is loaded into ultracold atoms, 
has been realized\cite{ultracold1,ultracold2,ultracold3}. 
In the system,
various parameters can be controlled experimentally such as
the hopping integrals, interaction strength, and lattice structure.
Furthermore, the spin and orbital degrees of freedom can be 
introduced~\cite{Li,Li2,Yb1,Yb2,Sr},
which makes strongly correlated fermion systems with multicomponents 
more interesting. 

One of the interesting and fundamental questions 
is how the symmetry breaking state 
is realized in the multicomponent system.
In our previous paper~\cite{Yanatori}, 
we have focused on the half-filled Hubbard model with repulsive interactions.
In the even component system, the magnetically ordered state,
where half of components occupy in the sublattice A and 
the others in the sublattice B, competes with
the correlated metallic state. When $N\ge 4$, the phase transition between
the correlated metallic and magnetically ordered states is of first order,
which is different from the second-order phase transition 
in the conventional SU(2) Hubbard model.
Furthermore, it has been clarified that the Mott transition 
without a magnetic instability indeed appears 
at finite temperatures when $N\ge 6$.
By contrast, it has been reported that, in the SU(3) system,
two ordered ground states are degenerate~\cite{Miyatake,Hasunuma}, and 
the phase transitions at finite temperatures are 
always of second order~\cite{YanatoriNCA,Yanatori}.

It has been found that in the repulsively interacting Hubbard model, 
the parity of the components plays an important role 
for ground state and finite temperature properties.
It has also been clarified that the maximum of 
the transition temperature is always lower than
that for the SU(2) system~\cite{Yanatori}. 
On the other hand,
in the SU$(N)$ attractive Hubbard model, 
which is one of the fundamental models,
a detailed analysis has been done only for 
$N\le 3$,\cite{Shiba,Scalettar,Freericks,Rapp1,Rapp2,IS1,IS2}
and
it has not been clarified systematically
how low temperature properties depend on the number of components.
Furthermore, it is naively expected that 
the density wave (DW) and/or superfluid (SF) states are more stable
than the magnetically ordered states in the repulsively interacting systems.
This may be important for the observations of the symmetry breaking state 
in the optical lattice~\cite{Hart}.
Therefore, it is desired to clarify
how stable the SF and DW states 
are against thermal fluctuations.

In the paper, 
we investigate the attractive SU$(N)$ Hubbard model in the infinite dimensions,
combining dynamical mean-field theory (DMFT)~\cite{DMFT1,DMFT2,DMFT3} 
with the continuous-time quantum Monte Carlo (CTQMC) 
method~\cite{Werner,solver_review}. 
We then discuss the stability of the spontaneously symmetry breaking states
at finite temperatures. 

The paper is organized as follows. In Sec. \ref{sec2}, 
we introduce the model Hamiltonian and briefly summarize our
theoretical approach. 
In Sec. \ref{sec3}, we discuss how the attractive
interaction realizes the SF and DW states 
at low temperatures.
A summary is given in the final section.

\section{Model and Method}\label{sec2}
We consider the multicomponent fermionic systems, which is described by 
the following SU($N$) Hubbard model,
\begin{eqnarray}
\hat{\cal{H}}=-t\sum_{\langle ij \rangle ,\alpha}
c^{\dagger}_{i\alpha}c_{j\alpha}
-U\sum_{i,(\alpha\beta)} 
\left(n_{i\alpha}-\frac{1}{2}\right)\left(n_{i\beta}-\frac{1}{2}\right),\label{model}
\end{eqnarray}
where $c^\dag_{i\alpha}$($c_{i\alpha}$) creates (annihilates) a fermion 
with color $\alpha(=1, 2,\cdots, N)$ at site $i$,
$\langle ij \rangle$ indicates the nearest neighbor sites,
$(\alpha\beta)$ indicates fermion pairs 
with different colors
and $n_{i\alpha}=c^\dag_{i\alpha} c_{i\alpha}$.
$t$ is the hopping integral and $U(>0)$ 
is the on-site attractive interaction 
between fermions with distinct components.
In the paper, we discuss the particle-hole symmetric systems with $\mu=0$.

The ground-state and finite-temperature properties 
in the two-component system ($N=2$)
have been discussed~\cite{Shiba,Scalettar,Freericks}.
In particular, in the infinite dimensions, the crossover between
the weak-coupling BCS state and strong-coupling BEC state, 
so-called BCS-BEC crossover,
has been discussed in much detail~\cite{Toschi,Bauer,Peters,Garg,KogaWerner}.
It is also known that 
the attractive Hubbard model is essentially the same as
the repulsive one under the particle-hole transformation~\cite{Shiba}.
Therefore, if the paramagnetic state is considered in the SU(2) system,
the increase of the interaction induces the phase transition in the system~\cite{Capone}.
In the paper, we refer the strong coupling state as the "Mott" state 
since its essential property is independent on the sign of 
the interaction.
As for the $N=3$ system, the competition between the SF and Mott states
(color SF and trion states) has been discussed~\cite{Rapp1,Rapp2,IS1,IS2}.
Namely, in the generalized SU$(N)$ attractive model, 
the Mott state should be regarded 
as the state where the composite particles and holons are randomly distributed.
It is instructive to examine the Mott transition 
since it plays an important role in the finite temperature
phase diagram in the repulsive SU$(N)$ model with $N>2$~\cite{Yanatori}.

In the paper, we consider the infinite dimensional Hubbard model 
to discuss how the SF and DW states are stable 
against thermal fluctuations.
To this end, we make use of DMFT~\cite{DMFT1,DMFT2,DMFT3},
in which the lattice model is mapped 
to the problem of a single-impurity connected 
dynamically to an effective medium.
The Green's function is obtained via the self-consistency conditions
imposed of the impurity problem.
The treatment is exact in $d\rightarrow \infty$ dimensions,
and even in three dimensions, DMFT has successfully explained
interesting physics in the solids and 
optical lattice systems~\cite{ultracold2}.
For simplicity, 
we use the fermion band by a semicircular density of state (DOS) 
$\rho(\epsilon)=2\sqrt{1-(\epsilon/D)^2}/(\pi D)$, 
which corresponds to an infinite-coordination Bethe lattice,
where $D$ is the half-bandwidth.

To discuss low temperature properties quantitatively,
we use the strong coupling version of 
the CTQMC method as an impurity solver~\cite{Werner,solver_review}.
This unbiased method can treat different energy scales correctly, 
which allows us to investigate the Hubbard model systematically.
To discuss the stability of the SF and DW states,
we calculate the pair potential between $\alpha$ and $\beta$ components
$\Delta_{\alpha\beta}$, and staggered moment $m$ as,
\begin{eqnarray}
\Delta_{\alpha\beta}&=&\frac{1}{L}\sum_i\langle c_{i\alpha}c_{i\beta}\rangle,\\
m&=&\frac{1}{NL}\sum_{i\alpha}(-1)^i\langle n_{i\alpha}\rangle-\frac{1}{2},
\end{eqnarray}
where $L$ is the number of sites.
To discuss the Mott transition in the strong coupling region,
we also calculate the quantity 
$z_\alpha=(1-{\rm Im} \Sigma_\alpha(i\omega_0)/\omega_0)^{-1}$
as a renormalization factor at finite temperatures.
The detail of DMFT is explicitly shown in Appendix~\ref{DMFT}.
In the weak coupling limit, 
it is hard to quantitatively deduce the critical temperature 
in terms of the CTQMC method. 
Here, we complementary use the static mean-field theory.
Its outline is given in Appendix~\ref{MFT}.

\section{Stability of the superfluid state}\label{sec3}

In the section, we consider the SU$(N)$ Hubbard model
at half filling to discuss the stability of the $s$-wave SF state.
In the state, fermion pairs are formed by two of $N$ components.
Therefore, in the system with odd $N$, 
fermions with $(N-1)$ components should form pairs and
unpaired fermions should yield metallic behavior in the system.
In fact, we find that one of the dispersions is gapless 
when the BCS theory with the pair potentials $\Delta_{\alpha\beta}$
is applied to the system (see Appendix~\ref{DMFT}).
Then, the energy spectrum inherent to the SF state 
should be given by
$
E_k=\sqrt{\epsilon_k^2+U^2\delta_2},
$
where $\delta_2=\sum |\Delta_{\alpha\beta}|^2/\lfloor N/2\rfloor$.
\begin{figure}[htb]
\centering \includegraphics[width=7cm]{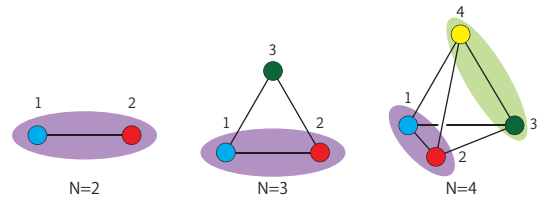}
\caption{
Fermion pairs in the multicomponent systems. 
Circles with the number represent the fermions with color, and
lines represent the interactions between fermions.
Shaded areas represent the fermion pairs discussed (see text).
}
\label{fig:pair}
\end{figure}
These should allow us to choose the simple pairing state to describe 
the SF state, as shown in Fig.~\ref{fig:pair}.
In the static mean-field theory, 
the interaction for fermion pairs with $\Delta_{\alpha\beta}$ stabilizes 
the SF state.
On the other hand, if the pair potential is not defined 
between certain components, 
the corresponding interaction simply shifts the energy and
plays no role in stabilizing the SF state (see Appendix~\ref{MFT}).
In the case, the gap equation for the SU$(N)$ attractive Hubbard model 
is reduced to that for the SU(2) case.
Thus, one expects that, in the SU$(N)$ model, 
the SF ground state is realized in the weak coupling region.
On the other hand, in the strong coupling region,  
the Mott state should be stabilized, where $N$-particle occupied sites and 
empty sites are equally realized.
Therefore, it is necessary to deal with strong correlations properly
beyond the simple BCS theory.
To examine low temperature properties in the Hubbard model systematically, 
we use DMFT, where the anomalous Green functions are introduced for
the corresponding pairs.

\begin{figure}[htb]
\centering
\includegraphics[width=8cm]{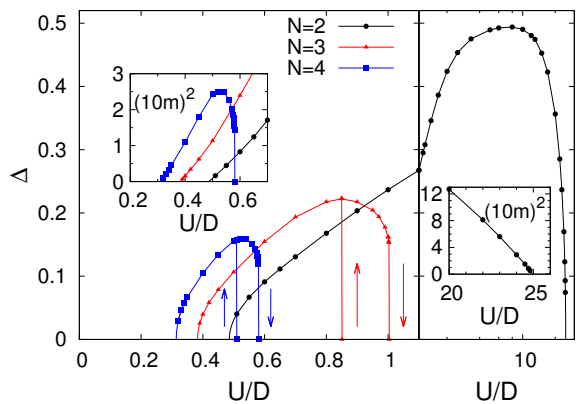}
\caption{
Circles, triangles, and squares represent pair potentials 
in the SU($N$) attractive Hubbard model at the temperature $T/D=0.01$.
The left (right) inset shows the critical behavior of the pair potentials
in the weak (strong) coupling region. 
Solid lines are guides to eyes.}
\label{fig:potential}
\end{figure}

By performing the DMFT calculations, we calculate the order parameters
for the SF state.
The results of the pair potential at the temperature $T/D=0.01$ 
are shown in Fig. \ref{fig:potential}.
Low temperature properties of the SU(2) system has been discussed 
in much detail~\cite{KogaWerner,Bauer,Garg,Peters,Toschi}.
In the weak coupling case ($U<U_{c1}$), 
a normal state is realized with $\Delta=0$.
The state has a Fermi surface for each color, which allows us to refer
it as the "metallic" state.
Beyond the critical value $U_{c1}$,
the attractive interaction leads to the formation of the Cooper pairs, and
the SF state is realized with the finite pair 
potential $\Delta(=\Delta_{12})$,
as shown in Fig.~\ref{fig:potential}.
Increasing the interaction, 
the pair potential has a maximum around $U/D\sim 8$ and decreases.
Finally, the pair potential vanishes at the critical value $U=U_{c2}$.
These critical interactions 
$U_{c1}/D\sim 0.49$ and $U_{c2}/D\sim 25$ are determined, 
by examining critical behavior $\Delta\sim |U-U_c|^\beta$
with the exponent $\beta=1/2$, 
as shown in the insets of Fig. \ref{fig:potential}.

\begin{figure}[htb]
\centering
\includegraphics[width=8cm]{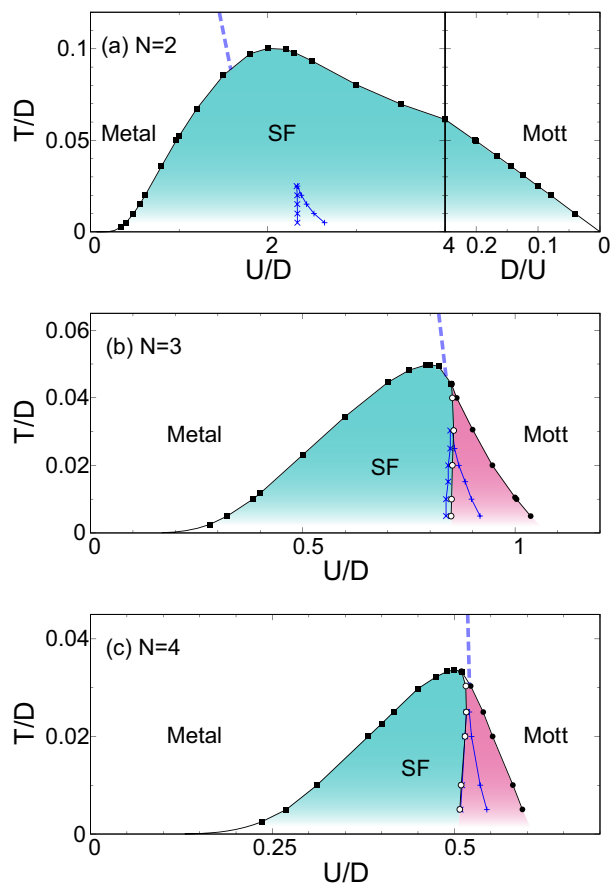}
\caption{
Finite temperature phase diagram 
for the SU$(N)$ attractive Hubbard model.
Solid squares represent the second-order SF 
phase transition points.
Open (solid) circles represent the transition
points, where the normal (SF) state disappears. 
Pluses (crosses) represent the
transition points under the paramagnetic condition, 
where the metallic (Mott) state disappears.
The crossover between the metallic and Mott states 
are shown as the dashed lines.
Solid lines are guides to eyes.
}
\label{fig:phase_SC}
\end{figure}

The larger $N$ leads to
different behavior in the strong coupling region.
It is found that the SF state is realized 
only in the narrower region $(U/D\lesssim 1)$,
as shown in Fig.~\ref{fig:potential}.
Furthermore, we clearly find the hysteresis in the order parameters.
This indicates that the first-order phase transition 
occurs between the SF and 
strong coupling Mott states.
The nature of the phase transition originates from 
the features for these two phases.
In the large $U$ region, fermions with $N$ components are tightly coupled to
each other. 
When we focus on the half-filled system, 
the composite particles and holons are mainly realized 
in the strong coupling region.
Therefore, this phase should be regarded as the "Mott" state.
On the other hand, the SF state is characterized by the fermion pairs.
The difference in their features yields the first order phase transition,
which is consistent with the results for $N=3$~\cite{Rapp1,Rapp2,IS1,IS2}.
This is essentially the same as the first-order phase transition 
discussed in the multicomponent fermion system 
with anisotropic repulsive 
interactions\cite{Inaba,Yanatori,Okanami,KogaWernerSC,HoshinoWerner}.
By contrast, the second-order phase transition occurs
in the weak coupling region.
The critical values are determined
as $U_{c1}/D\sim 0.38 (N=3)$ and $0.31 (N=4)$, by examining critical behavior
(see the left inset of Fig.~\ref{fig:potential}).
The finite temperature phase diagram is shown in Fig. \ref{fig:phase_SC}.
The SF state is widely realized in the SU(2) 
case~\cite{JarrellAF,GeorgesAF,KogaWerner}, 
while is realized in the narrower region with $U/D\lesssim 1$ 
for $N=3$~\cite{IS1} and 
$U/D\lesssim 0.7$ for $N=4$.
When $N>2$, the first-order phase transition occurs between the SF
and Mott states.

\begin{figure}[htb]
\centering
\includegraphics[width=8cm]{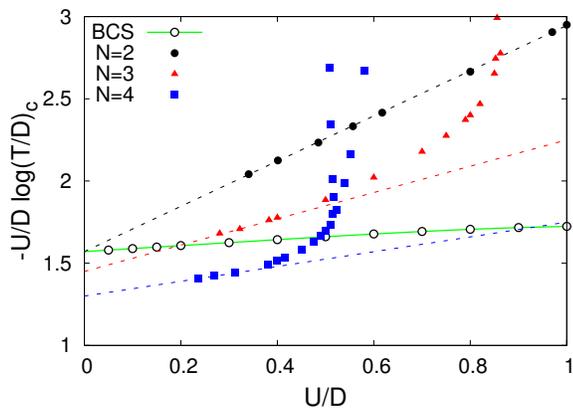}
\caption{
The quantity $-U/D \log (T/D)_c$ as a function of the interaction strength 
$U/D$ for the systems with $N=2, 3$ and 4.
Solid lines represent the results obtained by the static mean-field theory.
Dashed lines are guides to eyes.
}
\label{fig:detail_SC}
\end{figure}

If the system is paramagnetic,
the Mott transition is expected to occur between the metallic and Mott states
with increasing the interaction strength.
To discuss the role of the Mott transition in the attractive model,
we also calculate the renormalization factor $z_\alpha$.
At high temperatures, the curve of $z_\alpha$ has the inflection point, 
which means the crossover between metallic and Mott states.
On the other hand, at lower temperatures, the jump singularity appears instead,
suggesting the existence of the first-order phase transition.
Examining the singularity in the curve of $z_\alpha$,
we deduce the values $U_{c2}^{(1)}$ and $U_{c2}^{(2)}$,
where the Mott and metallic solutions disappear, respectively.
These boundaries are shown as the crosses and pluses 
in Fig.~\ref{fig:phase_SC}. 
It is found that, in the SU(2) system,
the Mott boundaries are much lower than the SF phase boundary 
obtained above.
By contrast, in the case with $N>2$, it is found that 
the Mott critical temperature, which is the end point of two Mott boundaries, 
is comparable with the maximum of the SF critical temperature.
This means that, in the $N>2$ case,
the SF state does not get the large energy gain 
in the strong coupling region,
which makes the SF state unstable and
yields the first-order phase transition to the Mott state.
On the other hand, in the weak coupling limit $U/D\sim 0$,
we find that the SF phase boundary 
$(T/D)_c\sim \exp(-\delta D/U)$ with a constant $\delta$.
This is qualitatively consistent with the conclusion of the BCS theory.
To clarify the $N$ dependence more clearly,
we show in Fig.~\ref{fig:detail_SC} the quantity $-U/D \log (T/D)_c$,
which reduces to $\delta$ in the weak coupling limit $(U\rightarrow 0)$. 
We also show the BCS results, shown as open circles 
in Fig. \ref{fig:detail_SC}.
When $U=0$, the DMFT results for $N=2$ are consistent with the BCS ones
although it is hard to extrapolate the value $\delta$ 
in the limit $U\rightarrow 0$.
This means that the BCS static mean-field theory captures 
SF fluctuations properly in the weak coupling limit.
The introduction of the interaction strength leads to
the slight difference between them.
In the larger $N$ cases, a discrepancy appears even in the weak coupling limit.
Therefore, in the $N>2$ case, it is necessary to take dynamical correlations
into account properly to discuss low temperature properties 
even in the weak coupling region.

In our discussions, the SF state is restricted to 
the simple pairing state, which should be
justified in the weak coupling limit. 
The generalized pairing state may be induced by strong correlations 
in the SU$(N)$ attractive Hubbard model, and
the phase boundary obtained above then gives us
a lower limit of the SF state.

\section{Stability of the density wave state}
\begin{figure}[htb]
\centering
\includegraphics[width=7cm]{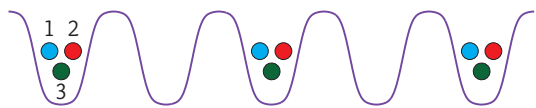}
\caption{DW state for the $N=3$ case.}
\label{fig:dw}
\end{figure}

\begin{figure}[htb]
\centering
\includegraphics[width=8cm]{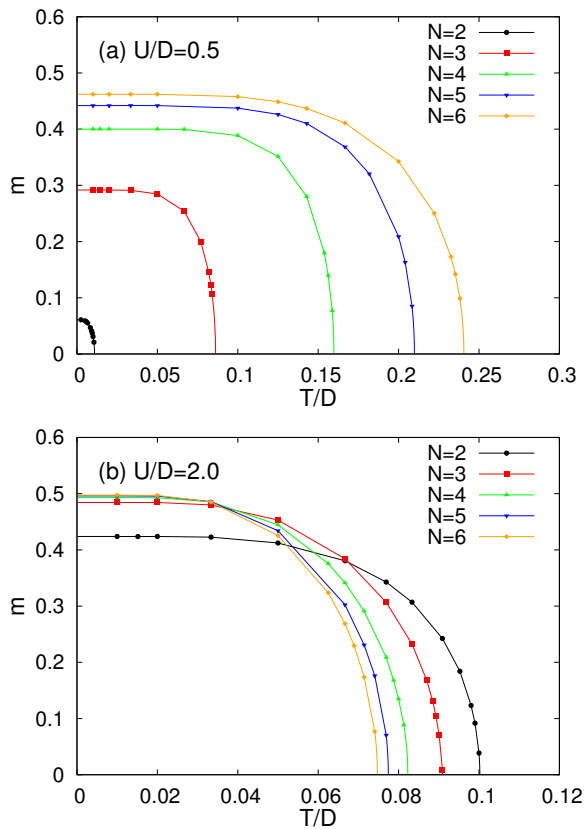}
\caption{Finite temperature phase diagram 
for the SU$(N)$ attractive Hubbard model.
Solid lines are guides to eyes.}
\label{fig:T}
\end{figure}

In the section, we consider the translational symmetry breaking state
in the bipartite lattice.
One of the most probable states is the DW state, where
$N$ fermions with distinct components are located in one of the sublattices,
and the empty state is realized in the other, 
as schematically shown in Fig.~\ref{fig:dw}.
Since this state is cooperatively stabilized 
by the attractive interactions,
it is expected to be more stable in the large $N$ case.
In fact, in the weak coupling region $U/D=0.5$, 
the order parameter $m$ monotonically increases with increase $N$ 
when the temperature is fixed, as shown in Fig.~\ref{fig:T}(a).
We also find the increase of the critical temperature.
On the other hand, when $U/D=2.0$, 
strong correlations play a crucial role in the finite temperature properties.
When one considers the second-order perturbation 
around the classical ground state, 
the energy scale characteristic of the DW state is given as
$\sim N/(N-1) D^2/U$.
Therefore, the critical temperature for the DW state
decreases with increase of $N$.
These are consistent with our numerical results shown in Fig.~\ref{fig:T}(b).

When the DW state is considered in the static mean-field theory, 
the SU$(N)$ system should be regarded as the SU(2) system with 
the renormalized interaction $(N-1)U$ (see Appendix \ref{MFT}).
Therefore, various quantities should be scaled in the weak coupling region.
\begin{figure}[htb]
\centering
\includegraphics[width=8cm]{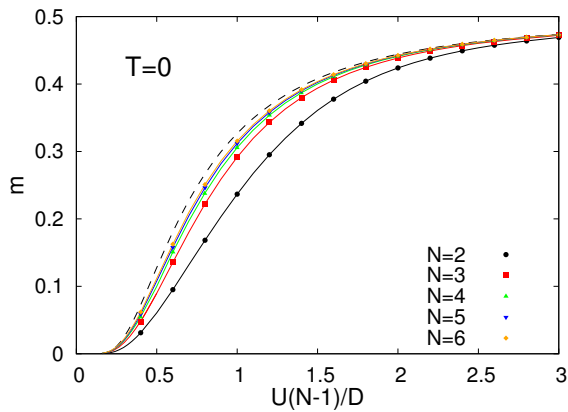}
\caption{
The staggered magnetization $m$ as a function of $(N-1)U$
in the systems with $N=2, 3, 4, 5$, and 6 at zero temperature.
Solid lines are guides to eyes.
}
\label{fig:zero}
\end{figure}
Figure \ref{fig:zero} shows the staggered magnetization at zero temperature,
by extrapolating data (see Fig.~\ref{fig:T}).
It is found that with increase of $N$, the staggered moments are well scaled, 
and converge to the mean-field results. 
This is consistent with the fact that the large value of $N$ suppresses 
quantum fluctuations and the conventional static mean-field theory
well describes ground state properties in the system.

\begin{figure}[htb]
\centering
\includegraphics[width=8cm]{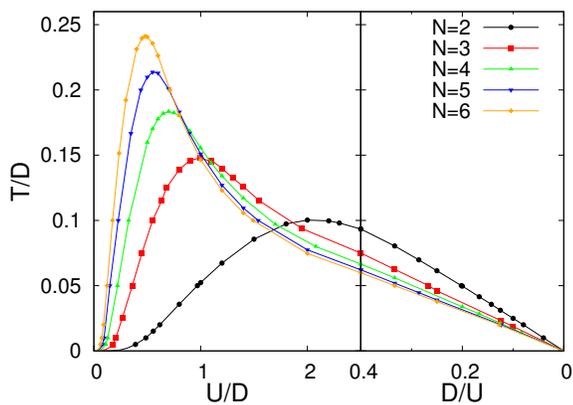}
\caption{Finite temperature phase diagram 
for the SU$(N)$ attractive Hubbard model.
Solid lines are guides to eyes.}
\label{fig:PD_at}
\end{figure}

Figure \ref{fig:PD_at} shows the phase diagrams 
for the DW state in the system with $N\le 6$.
It is known that, in the $N=2$ case, the SF and DW states 
are identical and thereby the phase diagrams 
are the same 
[see Fig. \ref{fig:phase_SC}(a)].
It is found that the phase diagrams for the DW state
are not changed qualitatively.
Namely, the second-order phase boundary in the strong coupling limit 
is well scaled by $ND^2/8(N-1)U$.
In the weak coupling limit, the phase boundary for the DW state
should be scaled by $(N-1)U$, contrast to that for the SF state. 
Therefore, we can say that the DW state is more stable than the SF state,
except for the $N=2$ case.

\begin{figure}[htb]
\centering
\includegraphics[width=8cm]{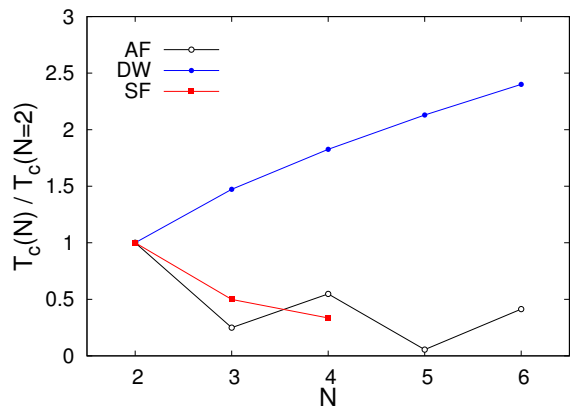}
\caption{
Solid circles and squares represent
the maximum critical temperatures for the DW and SF states
in the attractive Hubbard model.
The maximum critical temperatures for 
the antiferromagnetically ordered (AF) state 
in the repulsive Hubbard model~\cite{Yanatori} are shown as the open circles.
}
\label{fig:max}
\end{figure}

Figure~\ref{fig:max} shows the $N$ dependence of 
the maximum of the critical temperatures.
We find that with increase of $N$, 
the critical temperature for the DW (SF) state 
monotonically increases (decreases). 
This feature is in contrast to the magnetically ordered state 
in the repulsive Hubbard model~\cite{Yanatori}.
Therefore, in the attractive model, the parity of the components plays
no role for the nature of the phase transitions.
Since we have considered the Hubbard model in the infinite dimensions,
our system is far from realistic systems and 
we should not deduce the critical temperature quantitatively.
However, we believe that the ratio $T_c(N)/T_c(N=2)$ should be more reliable.
It is found that, in the $N\ge 5$ system, 
the maximum of the critical temperature 
for the DW state is 
more than twice higher than that for the $N=2$ system.
This temperature is accessible in the recent experiments~\cite{Hart}.
Therefore, the spontaneously translational symmetry breaking state 
should be observed
in the fermionic optical lattice system with large components
if the system is described by the SU$(N)$ attractive Hubbard model.

In our paper, we did not consider the condensation of the composite particles in the strong coupling region. It is naively expected that, in the odd $N$ case, the system can be regarded as free spinless fermions and the metallic ground state is realized. In the even $N$ case, the composite particles obey the Bose statistics, yielding the Bose-Einstein condensation. The characteristic temperature should be proportional to its effective hopping $D^N/U^{N-1}$ and the DW states should be much stable.

We also note that, in the bipartite system with $N>2$, 
the $s$-wave SF state has no chance to be realized
and the DW is widely realized instead.
If the system is doped, geometrically frustrated, and/or 
has three-body loss~\cite{Privitera,Titvinidze},
the DW state becomes unstable, which should induce
the phase separation, SF state, and supersolid state 
with both diagonal and off-diagonal order parameters.
Furthermore, in low dimensional systems, unconventional SF states
may be realized.
It is an interesting problem to clarify how these states 
compete with each other, which is now under consideration.

\section{Summary}

We have investigated spontaneously symmetry breaking states 
in the attractive SU($N$) Hubbard model at half filling.
Combining dynamical mean-field theory with the continuous-time 
quantum Monte Carlo method,
we have obtained the finite temperature phase diagrams 
for the SF and DW states.
As for the SF state, in $N>2$ case, 
it has been clarified that the second-order phase transition occurs 
in the weak coupling region, while
the first-order phase transition appears 
in the strong coupling region.
We have discussed low temperature properties for the DW state,
where the number of components plays a minor role.
We have clarified the monotonic increase of the maximum critical temperature
in the larger $N$ case.

\section*{Acknowledgments}
The authors would like to thank J. Nasu, K. Noda, 
and S. Suga for valuable discussions. 
Parts of the numerical calculations were performed
in the supercomputing systems in ISSP, the University of Tokyo.
This work was partly supported by the Grant-in-Aid for 
Scientific Research from JSPS, KAKENHI No. 25800193 and 
16H01066 (A.K.).
The simulations have been performed using some of 
the ALPS libraries~\cite{alps2}.

\appendix

\section{Dynamical mean field theory}\label{DMFT}
In this study, we have used DMFT to discuss the stability of the 
spontaneously symmetry breaking states in the attractive SU$(N)$ Hubbard model.
Here, we explain how to deal with the SF and DW states,
and the self-consistent equations are given. 
Solving the effective impurity problem 
by means of the CTQMC method based on the strong coupling expansion,
we have discussed low temperature properties in the SU$(N)$ Hubbard model.

\subsection{Superfluid state}
When the SF state is treated in DMFT,
the Green function should be described in the Nambu formalism~\cite{Georges}.
The matrix of the Green function depends on the formation of the Cooper pairs.
When the Cooper pairs are assumed as shown in Fig. \ref{fig:pair}, 
one introduces the simple Green function.
In the SU(4) case, it is explicitly given as,
\begin{equation}
\hat{G}(\tau)=\left(
\begin{array}{cccc}
G_1(\tau)&F_{12}(\tau)&0&0\\
F_{12}^*(\tau)&-G_2(-\tau)&0&0\\
0&0&G_3(\tau)&F_{34}(\tau)\\
0&0&F_{34}^*(\tau)&-G_4(-\tau)
\end{array} 
\right),
\end{equation}
where $G_\alpha(\tau)=-\langle T_\tau c_\alpha(\tau) c_\alpha^\dag(0)\rangle $
denotes the normal Green function, and 
$F_{\alpha\beta}(\tau)=-\langle T_\tau c_\alpha(\tau) c_\beta(0)\rangle$ and
$F^*_{\alpha\beta}(\tau)=-\langle T_\tau c_\beta^\dag(\tau) 
c_\alpha^\dag(0)\rangle$ denote the anomalous Green functions.
When the Bethe lattice is considered,
the self-consistency equation is given as
\begin{eqnarray}
{\cal \hat{G}}(i\omega_n)^{-1}=i\omega_n \hat{I}+\mu \hat{\Lambda}
-\left(\frac{D}{2}\right)^2 \hat{\Lambda} \hat{G}(i\omega_n) 
\hat{\Lambda},
\end{eqnarray}
where $\hat{I}$ is the identity matrix,
$\hat{\Lambda} = {\rm diag}(1,-1,1,-1)$,
and $\omega_n=(2n+1)\pi T$ is the Matsubara frequency.
Here, ${\cal{\hat{G}}}(i\omega_n)$ is the noninteracting Green's function 
for the effective impurity model.

When the SU(4) system is treated, one introduces two anomalous Green functions
$F_{12}(\tau)$ and $F_{34}(\tau)$.
In the case, the relative phase between two pair potentials 
$\Delta_{12}[=F_{12}(0_+)]$ and $\Delta_{34}[=F_{34}(0_+)]$ 
should play a minor role in stabilizing 
the SF state due to the high symmetry of the system.
In fact, we have confirmed that 
the magnitude of the pair potentials $|\Delta_{12}|$ and $|\Delta_{34}|$
does not change when these phases are parallel and antiparallel. 
If the system has an additional anisotopy such as the mixing between colors 
and/or the Hund coupling in the two-orbital model, 
the relative phase should be fixed.

\subsection{Density wave state}
We consider the DW state in the bipartite lattice, 
where the translational symmetry is spontaneously broken.
This diagonal order is simply described in terms of 
the Green's functions $G_{\alpha}^\gamma$ 
on the $\gamma(=A, B)$th sublattice.
The self-consistency equation is given as~\cite{Chitra}
\begin{eqnarray}
{\cal G}_\alpha^\gamma(i\omega_n)=i\omega_n+\mu-\left(\frac{D}{2}\right)^2
G_\alpha^{\bar{\gamma}}(i\omega_n),
\end{eqnarray}
where ${\cal{G}}_\alpha^\gamma(i\omega_n)$ 
is the noninteracting Green's function 
for the $\alpha$th component on the sublattice $\gamma$. 
Namely, the condition $G_\alpha^A(\tau)=G_\alpha^B(\beta-\tau)$ always 
appears at the symmetric case $(\mu=0)$.

\section{Static mean-field theory}\label{MFT}
In this study, we have used the static mean-field theory 
to discuss low temperature properties in the weak coupling region.

\subsection{Superfluid state}
The simple BCS theory is applied to our model Hamiltonian eq. (\ref{model}).
The interaction between $\alpha$ and $\beta$ components
should be decoupled as 
\begin{eqnarray}
Un_{i\alpha}n_{i\beta}\rightarrow U\Big(
\Delta_{\alpha\beta} c^\dag_{i\beta}c^\dag_{i\alpha}+
\Delta_{\alpha\beta}^* c_{i\alpha}c_{i\beta}
-|\Delta_{\alpha\beta}|^2
\Big),
\end{eqnarray}
where $\Delta_{\alpha\beta}=\langle c_{i\alpha}c_{i\beta}\rangle$ is 
the pair potential between $\alpha$ and $\beta$ components.
The mean-field Hamiltonian should be given as,
\begin{equation}
\begin{array}{rl}
H&\displaystyle =-t\sum_{\langle ij \rangle ,\alpha}
c^{\dagger}_{i\alpha}c_{j\alpha}
-U\sum_{i,\langle\langle \alpha\beta \rangle\rangle} 
\left(\Delta_{\alpha\beta} c^\dag_{i\beta}c^\dag_{i\alpha}+h.c.\right)
\\
&\displaystyle=\frac{1}{2}\sum_k\left(
\begin{array}{cc}
c_k&c_k^\dag
\end{array}
\right)\left(
\begin{array}{cc}
-\epsilon_k \hat{I}&U\hat{\Delta}\\
U\hat{\Delta}^+&\epsilon_k \hat{I}
\end{array}
\right)
\left(
\begin{array}{c}
(c_k^\dag)^t\\(c_k)^t
\end{array}
\right),
\end{array}
\end{equation}
where $c_k=(c_{k1}\;c_{k2}\;\cdots\;c_{kN})$, 
$c^\dag_k=(c^\dag_{k1}\;c^\dag_{k2}\;\cdots\;c^\dag_{kN})$,
$\epsilon_k$ is the dispersion for the fermions, 
$\hat{I}$ is an identity matrix, 
and $\hat{\Delta}$ is the alternative matrix with $\Delta_{\alpha\beta}$.
By using the Bogoljubov transformations, we obtain the Hamiltonian as,
\begin{eqnarray}
H&=&\sum_{k\alpha} E_{k\alpha} \gamma_{k\alpha}^\dag \gamma_{k\alpha}+E_0,\\
\frac{E_0}{L}&=&
\frac{1}{L}\sum_{k\alpha}\left(\epsilon_k
-E_{k\alpha}\right)+U\delta_2,
\end{eqnarray}
where $E_0$ is the ground state energy, 
$E_{k\alpha}$ is the dispersion relaion, 
$\delta_2=\sum_{(\alpha\beta)}|\Delta_{\alpha\beta}|^2/\lfloor N/2\rfloor$.
Since the pair potential is formed by two of $N$ components, 
one free fermion band appears in the system with the odd number of $N$.
The energy spectrum for the SF state is explicitly shown 
in the $N\le 4$ cases,
\begin{eqnarray}
E_k^{N=2,3}&=&\sqrt{\epsilon_k^2+U^2\delta_2}\\
E_k^{N=4}&=&\sqrt{\epsilon_k^2+U^2\left(\delta_2\pm
\sqrt{ \delta_2^2-|\delta_2'|^2 }\right)},
\end{eqnarray}
where $\delta_2'=\Delta_{12}\Delta_{34}-\Delta_{13}\Delta_{24}
+\Delta_{14}\Delta_{23}$, 
and $\Delta_{\alpha\beta}=\langle c_\alpha c_\beta\rangle$ 
is selfconsistently determined in terms of the generalized gap equations.
We have numerically confirmed that the condition $\delta_2^2=|\delta_2'|^2$
is satisfied in the converged solutions.
This allows us to choose the simple pairing states (see Fig.~\ref{fig:pair}) 
without loss of generality.
Then, we obtain the dispersion relations $E_k=\sqrt{\epsilon_k^2+U^2\delta_2}$
independent of the components $N$.

\subsection{Density wave state}
The DW state is characterized by the staggered moment $m$.
Then, the expectation value of the particle number should be given as,
\begin{eqnarray}
\langle n_{i\alpha}\rangle = (-1)^i m+\frac{1}{2}.
\end{eqnarray}
In the case, the interaction between $\alpha$ and $\beta$ components
should be decoupled as 
\begin{eqnarray}
&&U\left(n_{i\alpha}-\frac{1}{2}\right)
\left(n_{i\beta}-\frac{1}{2}\right)\nonumber\\
&&\rightarrow U\Big[
\left(n_{i\alpha}+n_{i\beta}-1\right)(-1)^im
-m^2 \Big].
\end{eqnarray}
Therefore, the mean-field Hamiltonian is given as,
\begin{equation}
H\displaystyle =-t\sum_{\langle ij \rangle ,\alpha}
c^{\dagger}_{i\alpha}c_{j\alpha}
-U(N-1)m\sum_{i\alpha} 
(-1)^i n_{i\alpha}.
\end{equation}
It is found that the above SU$(N)$ Hamiltonians are essentially the same as
the SU(2) one and
low temperature properties for the DW state 
should be scaled by $U\rightarrow U(N-1)$.


\end{document}